\documentclass[conference,anonymous=true]{IEEEtran}
\IEEEoverridecommandlockouts
\usepackage{cite}
\usepackage{amsmath,amssymb,amsfonts}
\usepackage{algorithmic}
\usepackage{graphicx}
\usepackage{textcomp}
\usepackage{xcolor}
\usepackage{paralist}
\usepackage{tabularx}
\usepackage{booktabs}
\usepackage{lmodern}
\usepackage{comment}
\usepackage{tikz}  

\newcommand{\circled}[1]{\tikz[baseline=(char.base)]{
            \node[shape=circle, draw, inner sep=1pt] (char) {#1};}}

\def\BibTeX{{\rm B\kern-.05em{\sc i\kern-.025em b}\kern-.08em
    T\kern-.1667em\lower.7ex\hbox{E}\kern-.125emX}}

\usepackage{float}
\usepackage{longtable}
\usepackage{caption}

\newif\ifhl{}
\hltrue{}
\hlfalse{}

\ifhl{}
 
\else
 
\fi

\newif\ifdraft{}
\drafttrue{}

\ifdraft{}
  \newcommand{\jhanote}[1]{ {\textcolor{purple} { ***[Shantenu]: #1 }}}
    
  \newcommand{\mtnote}[1]{ {\textcolor{orange} { ***[Matteo]: #1 }}}
  \newcommand{\ozgurnote}[1]{ {\textcolor{blue} { ***[Ozgur]: #1 }}}
  \newcommand{\miknote}[1]{ {\textcolor{brown} { ***[mikhail]: #1 }}}
  \newcommand{\NOTE}[1]{\phantom{}\begingroup\relax\ifmmode\boldmath\else\bfseries\fi\color{Cerulean}\ignorespaces#1\ignorespaces\endgroup}
  \newcommand{\TODO}[1]{\phantom{}\begingroup\relax\ifmmode\else\sffamily\fi\color{BurntOrange}\ignorespaces#1\ignorespaces\endgroup}
  \newcommand{\FIXME}[1]{\phantom{}\begingroup\relax\ifmmode\boldmath\else\bfseries\sffamily\fi\color{Red}\ignorespaces#1\ignorespaces\endgroup}
  \newcommand{\FIXED}[1]{\phantom{}\begingroup\relax\ifmmode\else\sffamily\fi\color{Green}\ignorespaces#1\ignorespaces\endgroup}
  \newcommand{\DELETE}[1]{\phantom{}\begingroup\relax\ifmmode\else\sffamily\fi\color{Red}\ifmmode\text{\sout{\ensuremath{#1}}}\else\sout{\ignorespaces#1\ignorespaces}\fi\endgroup}
\else
  \newcommand{\jhanote}[1]{}
  \newcommand{\mtnote}[1]{}
  \newcommand{\ozgurnote}[1]{}
  \newcommand{\miknote}[1]{}
  \newcommand{\NOTE}[1]{}
  \newcommand{\TODO}[1]{}
  \newcommand{\FIXME}[1]{}
  \newcommand{\FIXED}[1]{}
  \newcommand{\DELETE}[1]{}
\fi

\begin{document}

\pagestyle{plain} 


\title{Adaptive Protein Design Protocols and Middleware}

\newif\ifanon{}
\anontrue{}
\anonfalse{}
\ifanon{}
    \author{
    Anonymous Authors
    }
\else  

\author{%
  \IEEEauthorblockN{\small 
    Aymen Alsaadi\IEEEauthorrefmark{1}\textsuperscript{\textsection},
    Jonathan Ash\IEEEauthorrefmark{2}\textsuperscript{\textsection},
    Mikhail Titov\IEEEauthorrefmark{3},
    Matteo Turilli\IEEEauthorrefmark{1},
    Andre Merzky\IEEEauthorrefmark{4},
    Shantenu Jha\IEEEauthorrefmark{1}, and
    Sagar Khare\IEEEauthorrefmark{5}%
  }%
  \IEEEauthorblockA{\IEEEauthorrefmark{1} Department of Electrical and Computer Engineering, Rutgers University, Piscataway, NJ, USA}%
  \IEEEauthorblockA{\IEEEauthorrefmark{2} Institute for Quantitative Biomedicine, Rutgers University, Piscataway, NJ, USA}%
  \IEEEauthorblockA{\IEEEauthorrefmark{3} Brookhaven National Laboratory, Upton, NY, USA \\ 
  \texttt{mtitov@bnl.gov}}%
  \IEEEauthorblockA{\IEEEauthorrefmark{4} RADICAL-Computing Inc, Wilmington, DE, USA \\ 
  \texttt{andre.merzky@radical-computing.com}}%
  \IEEEauthorblockA{\IEEEauthorrefmark{5} Department of Chemistry and Chemical Biology, Rutgers University, Piscataway, NJ, USA}%
  \IEEEauthorblockA{Emails: \texttt{\{aymen.alsaadi, jonathan.ash, matteo.turilli, shantenu.jha, sagar.khare\}@rutgers.edu}}%
  \IEEEauthorblockA{\textsuperscript{\textsection} Equal contribution}%
}

\fi

\maketitle

\begin{abstract}

Computational protein design is experiencing a transformation driven by AI/ML. However, the range of potential protein sequences and structures is astronomically vast, even for moderately sized proteins. Hence, achieving convergence between generated and predicted structures demands substantial computational resources for sampling. The Integrated Machine-learning for Protein Structures at Scale (IMPRESS) offers methods and advanced computing systems for coupling AI to high-performance computing tasks, enabling the ability to evaluate the effectiveness of protein designs as they are developed, as well as the models and simulations used to generate data and train models. This paper introduces IMPRESS and demonstrates the development and implementation of an adaptive protein design protocol and its supporting computing infrastructure. This leads to increased consistency in the quality of protein design and enhanced throughput of protein design due to dynamic resource allocation and asynchronous workload execution. 
\end{abstract}


\section{Introduction}
Artificial intelligence and advances in computing have made it possible to design novel proteins tailored for specific purposes. However, the range of possible protein sequences and structures is astronomically vast, even for moderately long proteins. Therefore, achieving high convergence between generated and predicted structures demands significant computational resources for sampling. 

Combining AI systems with traditional HPC computations offers considerable scientific acceleration, measured by the number of high-quality structures achieved for a specific computational cost. The Integrated Machine Learning for Protein Structures at Scale (IMPRESS) framework enables the real-time coupling of AI systems with HPC tasks to improve protein design capabilities. In particular, IMPRESS speeds up the assessment of potential protein sequences compared to "vanilla” methods that do not utilize the real-time integration of AI and HPC capabilities. 

The integrated AI-HPC infrastructure and methodology enable the "evaluate as you go" approach to assess the effectiveness of models and evolve the specific computations used to generate data and train those models. The real-time coupling and concurrency of AI and HPC tasks are fundamental~\cite{brewer2024ai}, as they allow for bidirectional influence that enables AI systems to guide or inform HPC tasks and vice versa. For inverse problems like protein design, generative AI models provide an enormous space of solutions in which high-quality solutions must be efficiently generated at scale and validated as efficiently to identify promising designs. 

Methods requiring concurrent AI and HPC task execution necessitate advanced implementations that support real-time decisions on which tasks to execute and efficiently manage their execution. When coupled with traditional sequential execution models, inefficiencies like idle resources and prolonged workflow makespan arise. 

IMPRESS addresses these challenges by enabling adaptive decisions that influence the task set (i.e., the workload). It also supports asynchronous execution and dynamic resource allocation and management based on task demands, allowing concurrent task execution without dependencies causing idle waits. IMPRESS ensures that computational tasks can be adjusted in real time based on resource availability and task requirements, resulting in improved workflow and workload-level asynchronicity. By enhancing asynchronicity, IMPRESS reduces workflow makespan and boosts overall resource utilization. Finally, it facilitates the adaptive execution of heterogeneous workflows across diverse platforms. 

IMPRESS utilizes tools such as ProteinMPNN for the generation of sequences conditioned on protein backbones and AlphaFold for the structural prediction of candidate-designed proteins. As a demonstration model system, IMPRESS employs iterative runs of ProteinMPNN and backbone refinement techniques to optimize protein binder designs (PDZ domains) for specific peptide targets. PDZ domains are essential protein interaction modules, with broad binding specificity. They recognize the C-terminal 4-6 amino acid sequences of target proteins. Designing them for high affinity and selectivity for a particular C-terminus protein target is a key objective for drug and reagent development targeting various biological processes.

The main contributions of this paper are the development and implementation of an adaptive protein design protocol, along with its supporting computing infrastructure. This leads to increased consistency in the quality of protein design, as assessed by standard pLDDT, pAE and pTM metrics, and enhanced throughput of protein design. The improved throughput results from better utilization of computational resources, which, in turn, stems from dynamic resource allocation and asynchronous workload execution.

\section{IMPRESS Framework}
IMPRESS aims to enhance protein design by integrating AI-driven generative models with HPC simulations. This integration enables real-time feedback between AI and HPC tasks, thereby improving the design and production of proteins and, ultimately, foundational models through experimental data validation. IMPRESS increases the impact of AI/ML in protein design by developing and deploying sophisticated systems that facilitate the online integration of AI and HPC tasks. The combined AI-HPC framework and methodology allow for the real-time evaluation of model effectiveness and the adjustment of specific simulations used to generate data and train models. 

\subsection{Scientific Problem}
\label{subsec:science}
The computational design of proteins that bind to targets of interest has been a highly active area of research. This has been especially relevant in recent years, where the development of deep learning models such as AlphaFold2 and ProteinMPNN allows for the efficient and accurate determination of protein structure and sequence, respectively \cite{jepo21}\cite{dabo22}. These tools enable rapid generation and validation of binder designs to target a protein of interest, but the tolerated sequence space of the desired fold could span tens of thousands of unique samples, making evaluation difficult. To enable more efficient exploration of the protein sequence landscape, we introduce a genetic algorithm that couples AlphaFold2 and ProteinMPNN together to converge on optimal designs throughout several sequence generation and structure determination iterations.

\subsection{Design}
\label{subsec:design}

\begin{figure}
    \centering
    \includegraphics[width=0.48\textwidth]{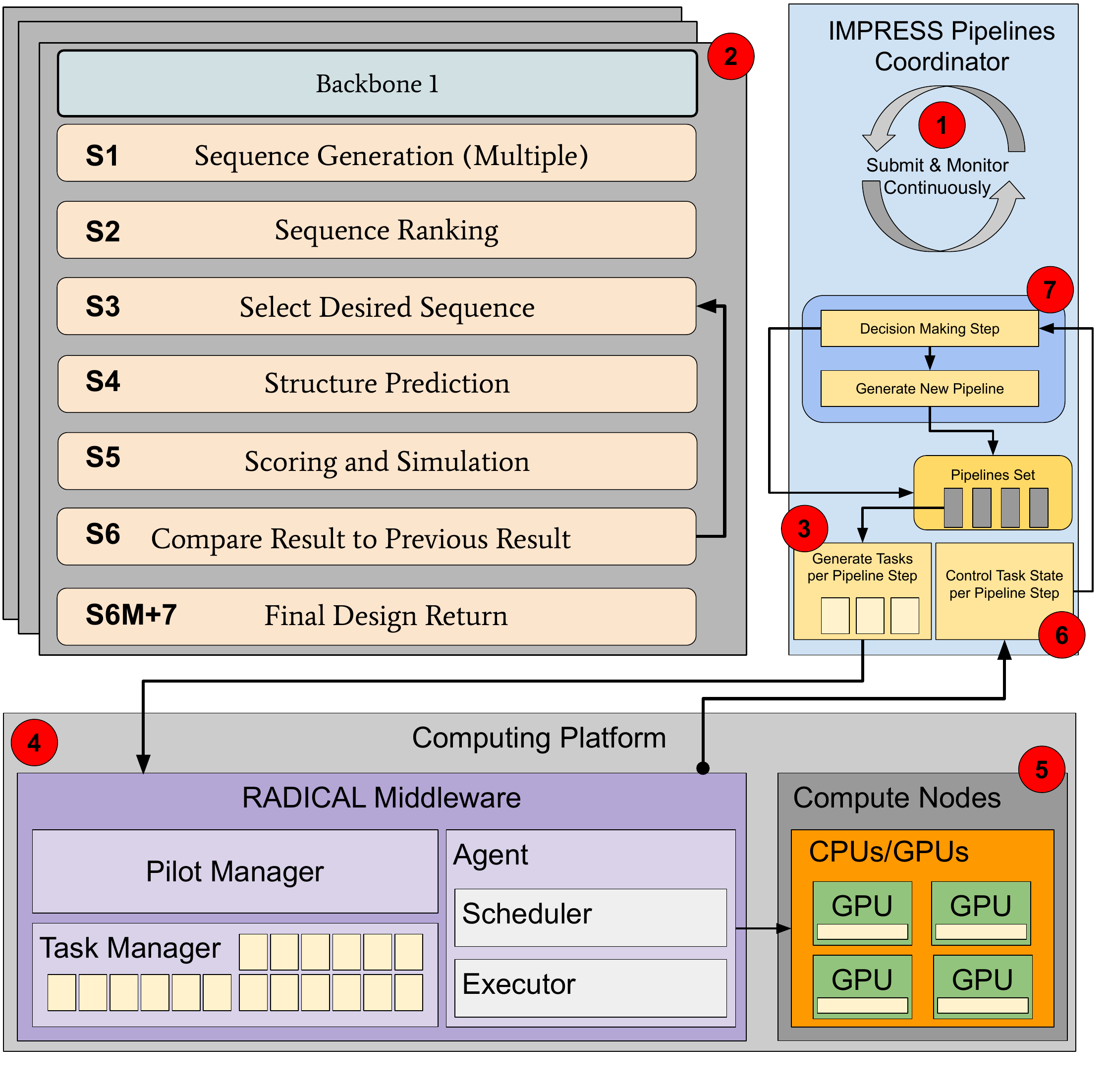}
    \vspace{-3ex}
    \caption{IMPRESS design and execution sequence}
    \label{fig:impress_design}
    \vspace{-3ex}
\end{figure}

Fig.~\ref{fig:impress_design} illustrates the design and execution sequence of the IMPRESS framework. The design follows a building block approach, where each component is a self-isolated, loosely coupled piece of code that can be extended, maintained, and replaced effortlessly without affecting other components~\cite{turilliblocks2019}.

IMPRESS consists of two key components: (i) a pipelines coordinator (refer to Fig.\ref{fig:impress_design} \circled{1}\circled{3}\circled{6}\circled{7}), and (ii) an execution runtime system represented by RADICAL-Pilot (RP)~\cite{merzkypilot2021}. We utilize RP to express IMPRESS pipelines and execute them on HPC resources (see Fig.\ref{fig:impress_design} \circled{4}\circled{5}). The coordinator 
manages the following processes iteratively:
(i) constructs and generates the IMPRESS pipelines, (ii) 
submits independent protein pipeline tasks concurrently for scheduling and execution based on resource availability~\cite{merzkypilot2021}, while tracking their execution states,
and (iii) makes adaptive decisions on submitting a new pipeline and with what characteristics. Specifically, the coordinator maintains a global perspective on each pipeline's results and the quality of the resulting sequences, which are later used to determine if there is a need to re-process ``low-quality'' sequences with a new pipeline.

 The proposed IMPRESS design features a closed-loop system that balances customization, iterative refinement, and automated quality control for improved protein engineering outcomes on HPC resources.

\subsection{IMPRESS Pipeline Structure}
\label{subsec:pipeline_strucuture}
The IMPRESS pipeline is a series of stages with one or more computing tasks.
It aims to optimize protein design through integrated sequence generation and structure prediction (see Fig.\ref{fig:impress_design} \circled{2}). 
\begin{inparaenum}[(i)]
    \item Stage 1 processes the input pipeline structures and generates 10 customizable sequences for each structure using ProteinMPNN, parameterized by user-defined settings (e.g., number of sequences, chains to design). 
    \item Stage 2 is the sequence selection process, which sorts the sequences from Stage 1 by their log-likelihood scores. 
    \item Stage 3 compiles the highest-ranking sequences into a fasta file for input into downstream tasks.
    \item Stage 4 employs AlphaFold to predict the structure from the fasta file. AlphaFold then ranks the candidate model structures by predicted TM-score (pTM), and returns the best complex.
    \item Stage 5 gathers quality metrics (pLDDT, pTM, inter-chain pAE) to assess iterative design improvements.
    \item Stage 6 compares the AlphaFold structure quality metrics from Stage 5 to previous iterations of the design generation. If the predicted structure confidence declines from the last iteration, Stages 4 and 5 are repeated with the next highest-ranked sequence by log-likelihood. This alternative selection process can be repeated up to 10 times, after which the pipeline is terminated. Alternatively, if the structure quality improves, the newly produced AlphaFold model serves as input into ProteinMPNN for the start of the next cycle.
    \item Stage 6M+7 represents the iterative cycling of the previous stages. After M repetitions, the final design candidates from the most recent cycle are returned to the user, along with all relevant quality metrics and statistics. 
 \end{inparaenum}

\subsection{Implementation}
\label{subsec:implementation}
We implemented the IMPRESS framework using the RADICAL-Pilot (RP) runtime system~\cite{balasubramanian2019radical}. RP is one component of RADICAL-Cybertools -- which are middleware building blocks that provide scalable, modular, and interoperable programming systems to execute heterogeneous workloads on heterogeneous HPC resources. RP supports different types of tasks, including OpenMP, MPI, and ML tasks, which are essential to enable the concurrent execution of AI and HPC tasks required by the IMPRESS framework.

RP API expresses the notion of tasks directly, which makes it flexible when building and managing adaptive workloads. RP does not provide an abstraction of a pipeline nor a workflow; thus, we implemented a~\texttt{Pipeline} class to bind a set of tasks that can be executed in a particular order and supported at runtime. 

The pipeline coordinator manages the concurrent and dynamic submission of pipelines using two communication channels: one to track new pipeline instances that need to be submitted to remote resources at the start or during the decision-making step and the other for completed tasks from each pipeline. IMPRESS operates in iterative stages during this implementation, submitting a single protein structure for each new pipeline.

The IMPRESS decision-making step determines the next steps by evaluating previous pipeline results, using ranking mechanisms based on their contribution to the overall structural prediction goals. It dynamically generates sub-pipelines when additional refinement, exploration, or iterative improvement is needed. These sub-pipelines are submitted for execution, enabling finer control over workflows. For example, if an initial task yields a coarse-grained prediction, a sub-pipeline is spawned to refine the resolution or explore alternative conformations.

\section{Evaluation}
We evaluate IMPRESS performance based on two key aspects: (i) computational performance on an HPC system and (ii) the scientific output regarding the protein quality produced. We compare the performance of IMPRESS adaptive pipelines implemented via RP, denoted as IM-RP, and the non-adaptive without RP, named control version, denoted as CONT-V, 
on HPC resources. We use three metrics: (i) AlphaFold confidence and error metrics (interchain pAE, pTM, and pLDDT), (ii) resource utilization for both CPUs and GPUs ($\%$), and (iii) the execution time of each implementation, which represents the total time taken by all tasks to finish the execution on the compute resources. 

Our experiments are conducted on Rutgers Amarel HPC~\cite{amarel2025} using one compute node with 4 Nvidia Quadro M6000 GPUs (12 GB each), 128 GB RAM, and 28 CPU cores. For both experiments highlighted above, we use the CONT-V as a baseline to compare IMPRESS framework results (IM-RP) scientifically and computationally. Table~\ref{table:exp-comparison} details the setup and results of our experiment.

\begin{table*}
    \caption{Experimental setup and results for CONT-V and IM-RP, concerning computational and scientific metrics. IM-RP demonstrates overall improvements in both computational efficiency and the quality of the produced protein. PL=Pipeline}
    \label{table:exp-comparison}
    \centering
    \resizebox{\textwidth}{!}{ 
    \begin{tabular}{lcccccccccc} 
        \toprule
        Approach & \# PL & \# Sub-PL & \# Structures/PL & Trajectories &CPU \% & GPUs \%  & Time (h) & pTM Net $\Delta$ (\%) & pLDDT Net $\Delta$ (\%) & pAE Net $\Delta$ (\%) \\
        \midrule
        \midrule
        CONT-V  & 1  & N/A & 4 & 16 & 18.3\% & 1\% & 27.7  & 0.28 (--) & 5.8 (--) & -6.7 (--) \\
        \midrule
        IM-RP   & 2  & 7   & 4 & 23 & 88\% & 61\% & 38.3  & 0.32 (\textbf{+14.3\%}) & 7.7 (\textbf{+32.8\%}) & -6.61 (\textbf{+1.3\%}) \\
        \bottomrule
    \end{tabular}
    } 
    \vspace{-2ex}
\end{table*}

\subsection{The Effect of Adaptivity on Optimal Protein Sequences}
\label{eval:subsec:science}
We implemented a simple testing protocol to verify that our genetic algorithm can outperform traditional iterative design methodologies. We prepared structures for four unique PDZ Domains (NHERF3, HTRA1, SCRIB, and SHANK1) and placed them in a complex with the last 10 residues of Alpha Synuclein. The IM-RP pipeline optimized these PDZ Domains for four design cycles, gradually enhancing their interactions with the target peptide. As a basis for comparison, we also prepared a control pipeline (CONT-V), which consists of all the IM-RP stages but lacks adaptive decision-making between cycles. The starting structures for the two pipelines were kept the same. Ten sequences for each complex were generated with ProteinMPNN at the beginning of the CONT-V implementation. One was chosen randomly to have its structure predicted with AlphaFold. The new structure was fed into ProteinMPNN for the next cycle. Performance was not compared between iterations, and trajectories were not pruned if they deteriorated. This process was repeated 4 times, and the AlphaFold metrics collected at each iteration were compared to the structure statistics obtained from the original IM-RP methodology tasked with the same design problem.

\begin{figure}
    \centering
    \includegraphics[width=0.48\textwidth]{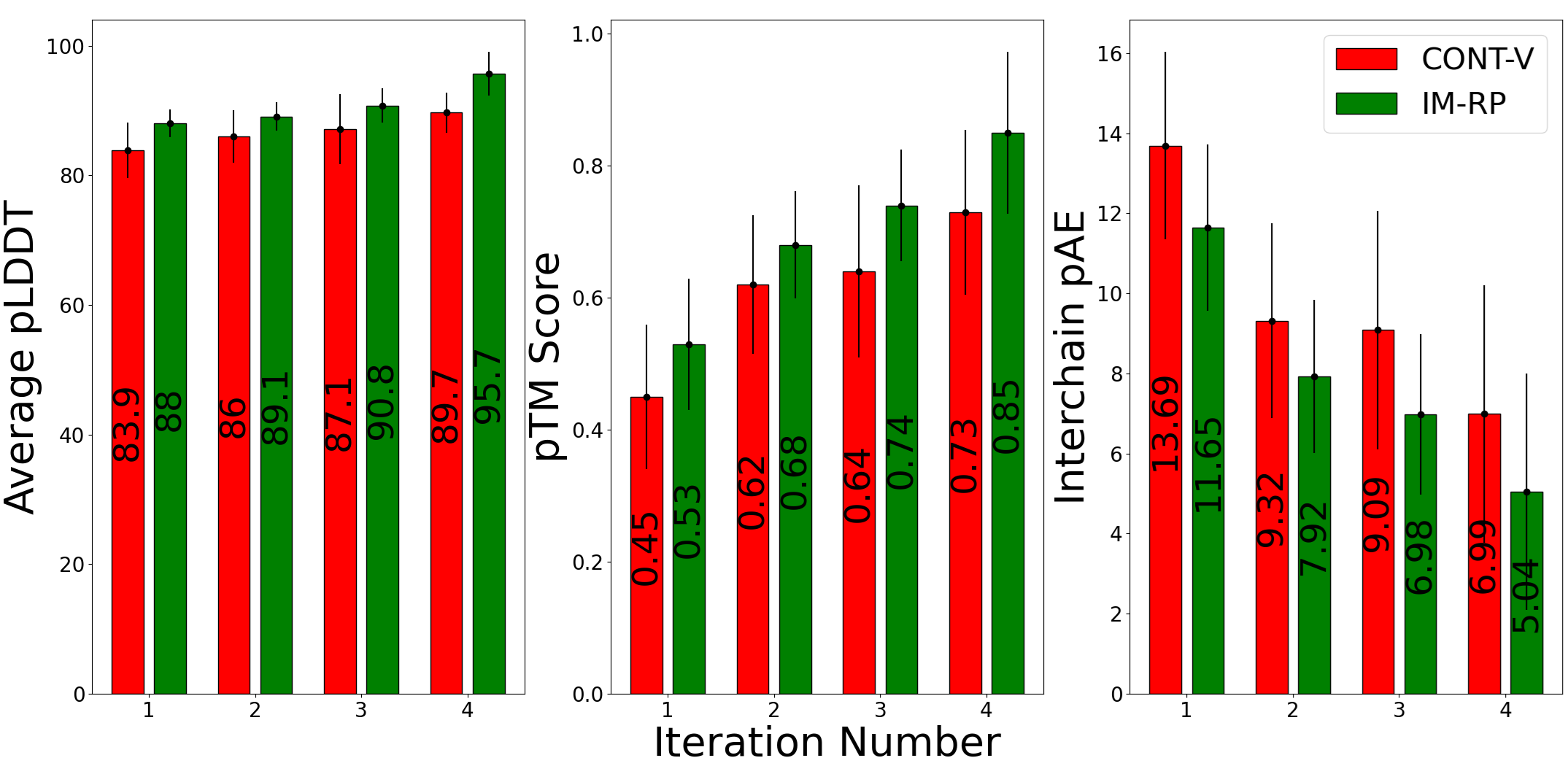}
    \caption{Comparison of AlphaFold pLDDT (Left; higher is better), pTM (Center; higher is better), and Interchain pAE (Right; lower is better) between CONT-V and IM-RP pipelines. Bars show median values for each metric across 4 PDZ-peptide structures, with CONT-V in red and IM-RP in green. Error bars represent half a standard deviation.}
    \label{fig:scientific_metrics_vs_control}
    \vspace{-3ex}
\end{figure}

As shown in Fig. \ref{fig:scientific_metrics_vs_control}, while the CONT-V mechanism gradually improved the design set with each iteration, the IM-RP pipeline attained superior results. The genetic mechanism enabled IMPRESS to achieve higher pLDDT, higher pTM, and lower inter-chain pAE medians compared to CONT-V at every iteration. Furthermore, the adaptive protocol demonstrated a higher consistency in design quality, as indicated by the lower standard deviation in the pLDDT and pTM metrics. While TABLE \ref{table:exp-comparison} shows that the IM-RP implementation took longer to complete than CONT-V, this is because IM-RP evaluated more trajectories for the adaptive decision making process. In total, while CONT-V only examined 16 trajectories from the starting structures, IM-RP evaluated 23 unique trajectories across 7 separate sub-pipelines.

\begin{figure}
    \centering
    \includegraphics[width=0.48\textwidth]{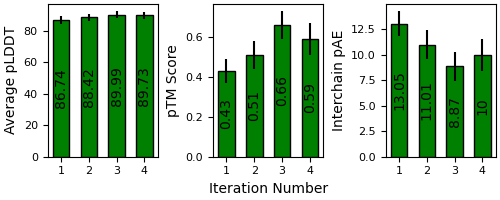}
    \caption{Achieved AlphaFold pLDDT (Left; higher is better), pTM (Center; higher is better), and Interchain pAE (Right; lower is better) by the expanded IM-RP workflow. Bars show median values for each metric across 70 PDZ-peptide structures. Error bars represent half a standard deviation.}
    \label{fig:impress_all}
    \vspace{-3ex}
\end{figure}

We further demonstrated the capabilities of the IM-RP framework by expanding our design pipelines to 70 unique, experimentally resolved PDZ-peptide complexes mined from the Protein Data Bank. These pipelines were tasked with a similar objective, where all PDZ domains were placed in complex with the last four residues of Alpha Synuclein, then iteratively optimized over the course of four design cycles to gradually improve their interactions with the target. In total, IM-RP examined 354 trajectories across 96 unique sub-pipelines. As shown in Fig. \ref{fig:impress_all}, all AlphaFold metrics improved continuously during the first three iterations. The median quality of the fourth iteration deteriorated, as adaptivity was not enforced in the final design cycle. The clear drop in design quality demonstrates the importance of our selection criteria, as the pipelines failed to resume established positive metric trends in its absence. 

\subsection{Heterogeneous Computational Resources Utilization}
\begin{figure}
    \centering
    \includegraphics[width=0.48\textwidth]{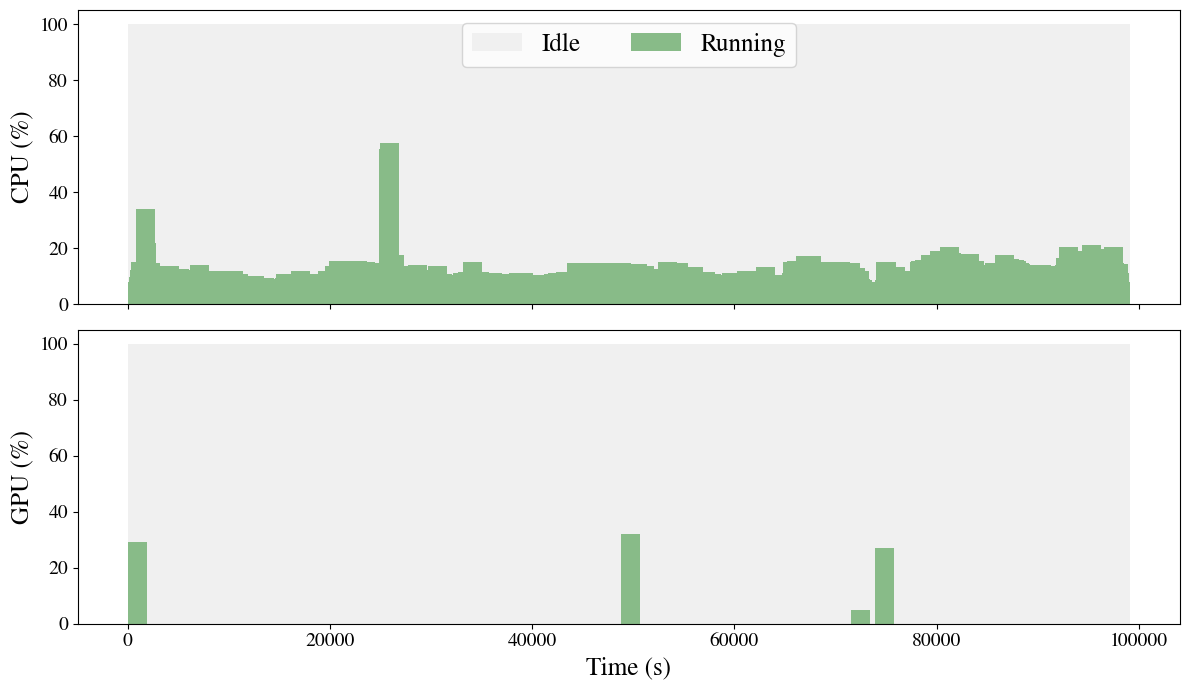}
    \caption{CONT-V total GPU/CPU resource utilization and execution time.}
    \label{fig:res_util_control}
    \vspace{-3ex}
\end{figure}

Fig.~\ref{fig:res_util_control} shows the total resource of CONT-V on 28 CPU cores (top) and 4 GPUs (bottom). The average CPU utilization is $\sim$18.3\%, while GPU utilization is much lower at $\sim$1\%, indicating that only one GPU was used. This under-utilization occurs because CONT-V deploys four structures for each ProteinMPNN call sequentially. Additionally, the AlphaFold construction phase runs on CPU, which takes hours to finish due to large databases and I/O bottlenecks, while GPUs remain idle, resulting in low GPU utilization~\cite{parafoldbozitao2022}.

\begin{figure}
    \centering
    \includegraphics[width=0.48\textwidth]{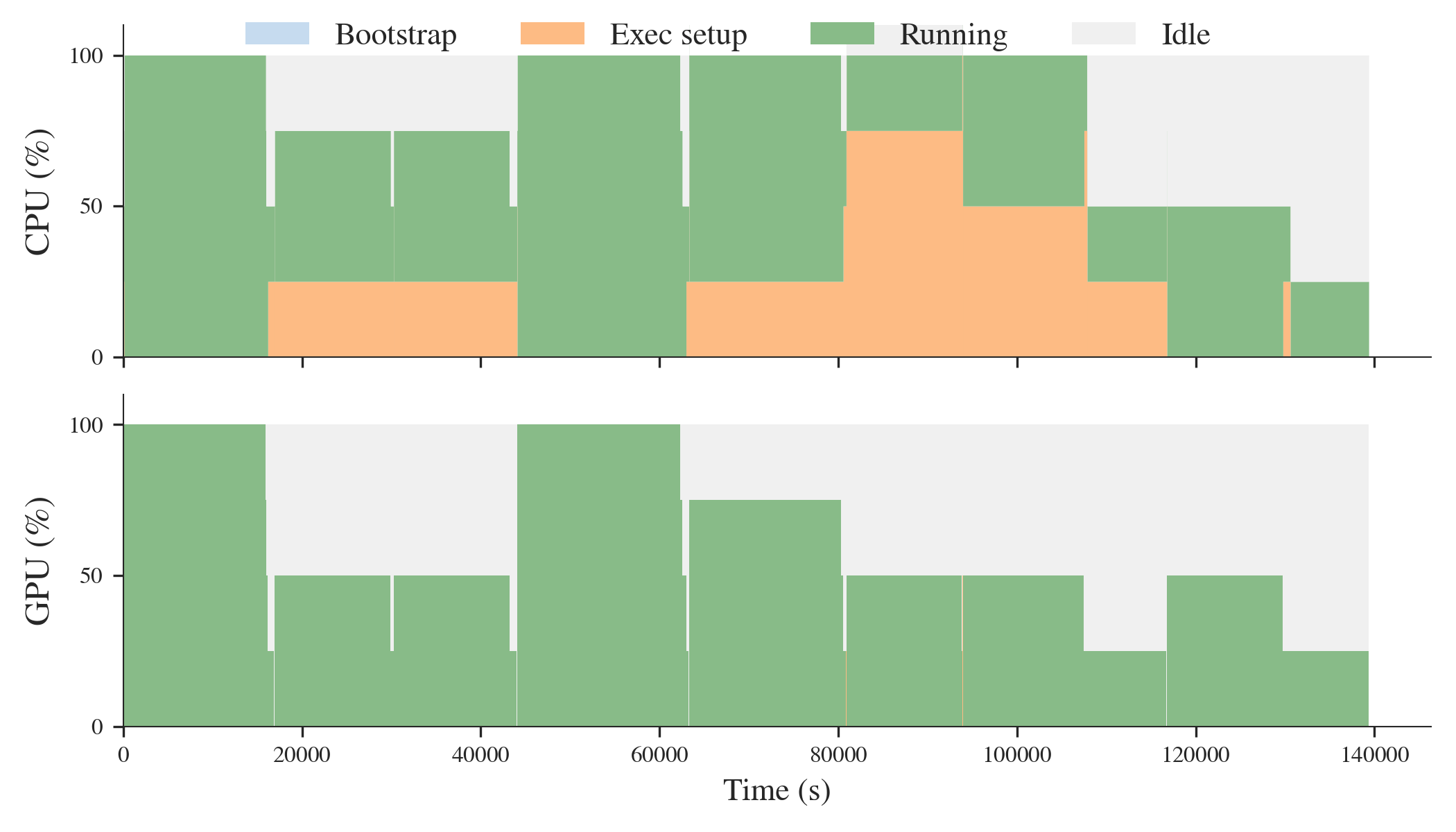}
    \caption{IM-RP total GPU/CPU utilization and execution time. Bootstrap: RP startup time. Exec setup: time for RP to prepare task execution (including script creation and sandbox setup; time varies depending on the file system). Running: task execution time on assigned resources.}
    \vspace{-3ex}
    \label{fig:res_util_impress}
\end{figure}

Fig.~\ref{fig:res_util_impress} shows the total resources the IM-RP implementation utilizes of CPUs (top) and GPUs (bottom). The results show that IM-RP outperforms the CONT-V as it highly utilizes the available CPUs ($\sim$88\%) and GPUs ($\sim$61\%). The IM-RP adaptive design smartly takes advantage of the available resources by offloading the newly created pipelines during the decision-making stage (see Fig.~\ref{fig:impress_design}) to the idle resources when possible.

Fig.~\ref{fig:res_util_control} and Fig.~\ref{fig:res_util_impress} show the execution time of CONT-V and IM-RP, respectively. CONT-V offers lower execution time due to processing less trajectories (see Table.~\ref{table:exp-comparison} and~\ref{eval:subsec:science}).

This behavior hinders resource utilization as less work is processed, and thus, fewer resources are utilized. More importantly, the quality of the produced design by CONT-V is lower because the fixed-structure approach limits exploration of the design space, reducing the opportunity to identify optimal configurations.

\section{Related Work}
With the advent of deep learning models that enable high-level insight into protein sequence and structure, several other advances have been made in developing genetic algorithms for design generation.

The recently proposed EvoPro protocol achieves this through iterative runs of sequence generation—using either ProteinMPNN or random mutagenesis—and AlphaFold, which employs single-sequence structure prediction mode to accelerate inference\cite{gnkrk23}. 
EvoPro shows significant advancements in coupling deep learning models for efficient sequence space exploration but faces two limitations. Eliminating MSA generation restricts EvoPro to designing binders of a specific size\cite{apco21}. As the designed domain increases in complexity, AlphaFold2 needs to utilize evolutionary information to ensure an accurately predicted structure\cite{e23}. Furthermore, AlphaFold2 is a powerful classifier, distinguishing between high- and low-quality binders\cite{mdbabb23,dcso24}. Allowing AlphaFold2 to utilize evolutionary information in its constructed MSA improves its predictive abilities and directs IMPRESS toward effectively guiding design exploration for more potent binders.

The newly developed MProt-DPO protocol is another technique that enables efficient exploration of the protein design landscape \cite{dhbo24}. Sequences are generated with a protein LLM, then ranked and sorted into preference pairs to fine-tune the original model \cite{laro23}\cite{rsmo24}.

MProt-DPO achieves remarkable efficiency but is restricted to purely sequence-based models. Therefore, design generation is never directly conditioned on the protein structure, but must rely on downstream MD simulations.

The IMPRESS framework allows any sequence generation method to be plugged into the design pipeline, enabling both LLMs and graph-based models to fully exploit the rich functional information available in protein structures. 

\section{Conclusion \& Future Work}
We introduced IMPRESS, a framework capable of continuously coordinating protein design pipelines that require the adaptive execution of AI and HPC tasks concurrently. IMPRESS achieves better results than traditional approaches, and its modular design makes it agnostic to the type of scientific problem.

The pipeline can run ProteinMPNN on a two-chain complex to redesign the entire receptor, creating new and improved interactions with the substrate. Through several iterations of sequence generation with ProteinMPNN and structure determination with AlphaFold, IM-RP continuously enhances its design pool, eventually converging on highly optimized candidates that exhibit stronger interactions with the target peptide. IM-RP is an adaptive workflow capable of pruning design trajectories if their quality declines compared to previous iterations. With this approach, IM-RP facilitates highly efficient exploration of the design landscape and yields higher quality binders than traditional sequential design methodologies.

This characteristic is further illustrated in our evaluation protocol. We have applied this pipeline to various PDZ domains to engineer new Alpha Synuclein binders. In doing so, we have demonstrated that IM-RP performs better than the CONT-V methodology for the same design challenge. This is evident in the IM-RP design set's higher confidence scores and its more efficient use of compute resources compared to the CONT-V implementation. 

Our next goal is to generalize this pipeline, applying iterative sequence generation and structure prediction rounds to proteases to improve catalytic activity. To accomplish this, ProteinMPNN runs must fix the catalytic residues rather than design the entire protein. Furthermore, as AlphaFold has difficulty accurately placing the peptide in protease complexes, we will instead predict our designs in monomeric form. The pipeline will be demonstrably generalizable across various protein design problems by implementing these simple changes.

We are enhancing IMPRESS to support the real-time evaluation and optimization of foundation models and evolve the specific simulations used to generate data and train models. IMPRESS will provide a scalable and generalized computational platform that will use integrated AI-HPC computing for the efficient generation of
high-quality proteins while enhancing foundational models using real-world experimental data.

{\footnotesize {{\bf Acknowledgements} NSF-2438557, 2103986 and 1931512; Rutgers-New Brunswick Cyberinfrastructure and AI for Science and Society}

\vspace{-0.1in}

\bibliographystyle{IEEEtran}
\bibliography{main}

\end{document}